\documentstyle[multicol,aps,prl,epsf]{revtex}

\begin{document}

\draft

\title{Accumulation and depletion layer thicknesses in organic field effect transistors}

\author{Manabu Kiguchi$^1$, Manabu Nakayama$^1$, Kohei Fujiwara$^1$,
Keiji Ueno$^2$, Toshihiro Shimada$^3$ and Koichiro Saiki$^1,3$}

\address{$^1$Department of Complexity Science $\&$ Engineering, Graduate
School of Frontier Sciences, The University of Tokyo, \\7-3-1 Hongo,
Bunkyo-ku,Tokyo 113-0033, Japan}
\address{$^2$Department of Chemistry, Faculty of Science,
Saitama University, Saitama, Saitama 338-8570, Japan}
\address{$^3$Department of Chemistry, Graduate School of Science, The University of Tokyo, Bunkyo-ku, Tokyo 113-0033, Japan}

\date{\today}

\maketitle

\begin{abstract}
We present a simple but powerful method to determine the thicknesses of
 the accumulation and depletion layers and the distribution curve of
 injected carriers in organic field effect transistors. The conductivity
 of organic semiconductors in thin film transistors was measured in-situ
 and continuously with a bottom contact configuration, as a function of
 film thickness at various gate voltages. Using this method, the
 thicknesses of the accumulation and depletion layers of pentacene were
 determined to be 0.9 nm ($V_{G}$=-15 V) and 5 nm ($V_{G}$=15 V).
\end{abstract}

\medskip

\begin{multicols}{2}
\narrowtext

\section{INTRODUCTION}
\label{sec1}
Electric-field control of physical properties is highly desirable from
fundamental and technological viewpoints, because it does not introduce
any chemical or microscopic structural disorder in the pristine
material\cite{1,2}. This is also a basis of field effect transistors
(FET), in which accumulation, depletion, and inversion layers are formed
at the interface\cite{3}. "Where is the region, in which carriers are
practically injected or depleted by electric fields?" is, thus, an
important problem relating closely with the operation principle of FET
and the electric-field control of physical properties. Although it is
quite a fundamental issue, there had been no investigation on evaluating
experimentally the dimension of these regions.
 
\begin{figure}
\begin{center}
\leavevmode\epsfysize=30mm \epsfbox{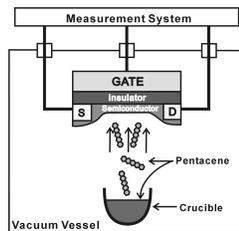}
\caption{Schematic experimental layout. The source(S)-drain(D) current was measured in-situ at various gate voltages.}
\label{fig1}
\end{center}
\end{figure}

We propose a simple but powerful method to observe directly the regions
in which the carriers are exhausted or injected by electric
fields\cite{4}. As shown in the Fig.~\ref{fig1}, the source(S)-drain(D)
current in thin film transistors (TFT) is measured as a function of the
film thickness with a bottom-contact configuration. The gradual change
in the conductivity is measured in-situ under the same condition at
different gate voltages ($V_{G}$) avoiding the problem of specimen
dependence which is often encountered in the organic films due to the
difference in grain size, crystallinity, interface state, etc. This
in-situ and continuous measurement would provide information on the
accumulation and depletion layers, which will be described later.
 
Despite the above mentioned advantage, the in-situ and continuous
measurement had not been examined except for a few cases\cite{5}. The
major reason is that higher substrate temperature is required to obtain
a flat film for inorganic semiconductors, and hence the measurement is
difficult to be applied. In contrast, a flat organic film can be
obtained even at low substrate temperature, which enables the in-situ
and continuous measurement during growth of organic films.

In the present study, we applied the measurement to pentacene, and
evaluated the thickness of the depletion layer and distribution curve of
injected carriers in accumulation layer.
  
\section{EXPERIMENTAL}
\label{sec2}
The substrate is a highly doped silicon wafer, acting as a gate
electrode. The gate dielectric layer is a 700 nm thermally grown silicon
dioxide. On top of the surface, 30 nm thick gold S, D electrodes were
deposited through a shadow mask. The channel length and the channel
width were 100 $\mu$m and 5.4 mm. The organic materials used were
pentacene (Aldrich), sexithiophene ($\alpha$-6T; synthesized by Syncom
BV), and C$_{60}$ (Materials and Electrochemical Research). Organic
films were deposited by means of vacuum deposition. The substrate
temperature was kept at 310 K during the growth. All the measurements
were performed under high-vacuum condition. 

\section{RESULT and Discussion}
\label{sec3}

\begin{figure}
\begin{center}
\leavevmode\epsfysize=40mm \epsfbox{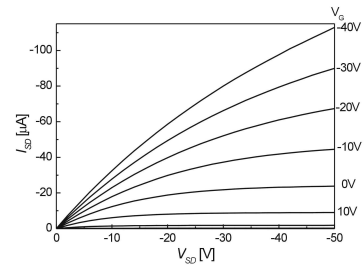}
\caption{Source (S)-drain (D) current ($I_{SD}$) versus S-D voltage ($V_{SD}$) characteristic of pentacene TFT at various gate voltages ($V_{G}$).}
\label{fig2}
\end{center}
\end{figure}

Figure~\ref{fig2} shows S-D current ($I_{SD}$) versus S-D voltage
($V_{SD}$) for the 20 nm thick pentacene TFT. Typical p-type
semiconductor features are exhibited. We observed ohmic behavior of the
S, D contacts to the pentacene active layer. The carrier mobility,
estimated in the linear transport regime, was 0.23 $cm^{2}V^{-1}s^{-1}$,
which is comparable to the world's highest TFT mobility value\cite{6}.
 
\begin{figure}
\begin{center}
\leavevmode\epsfysize=50mm \epsfbox{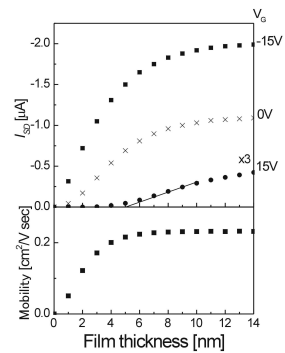}
\caption{a) Source-drain current ($I_{SD}$) of pentacene measured as a
 function of film thickness at various gate voltages ($V_{G}$);
 15(circle), 0(cross), and -15 V(square). Source-drain voltage was kept
 at 1 V. b) Mobility as a function of film thickness. }

\label{fig3}
\end{center}
\end{figure}

We then discuss the depletion layer thickness from the in-situ and
continuous measurement. $V_{SD}$ was kept at 1 V and $V_{G}$ was set at
three different voltages; +15, 0, -15 V. Electric fields were not
applied during the growth in order to exclude the influence of the
electric current on the film growth. Figure~\ref{fig3} shows $I_{SD}$ as
a function of pentacene film thickness at different $V_{G}$. TFT
characteristics were measured at each film thickness to obtain the
mobility, and the thickness dependence of the mobility was included in
the bottom part of Fig.~\ref{fig3}. Here we find the presence of
threshold thickness ($d_{th}$) at which electric current begins to
flow. There are two points to note about $d_{th}$. 
First, $d_{th}$ is 0.6 nm
at $V_{G}$=0 V, and electric current is observed even for the 1.0 nm
thick film. Considering the thickness (1.5 nm) of 1 monolayer (ML)
pentacene, the present result indicates that the S, D electrodes (width
100 $\mu$m) can be electrically connected by only 1 ML
pentacene\cite{7}. 
Second, $d_{th}$ shows clear $V_{G}$ dependence, and it shifts
up to 5.0 nm at $V_{G}$=15 V. When positive $V_{G}$ is applied to p-type
semiconductors, the depletion layer with a low conductance is formed at
the semiconductor/insulator interface. Therefore, on the thin limit, the
whole film can be depleted, and $I_{SD}$ does not flow at positive
$V_{G}$. 
Consequently, $d_{th}$ (5.0 nm) corresponds to the thickness of the
depletion layer at $V_{G}$=15 V. By the in-situ and continuous
measurement, the thickness of the depletion layer can be observed
directly as a shift of $d_{th}$.
 
In order to discuss the meaning of the depletion thickness, we will try
to apply the conventional model which has been already established for
inorganic semiconductors. In the depletion approximation\cite{3,8}, the
carrier density is given by $N=Q/eT_{s}$ where
$Q=\epsilon_{ox}V_{G}/T_{ox}$, $T_{s}$: thickness of the depletion
layer, $\epsilon_{ox}$: dielectric constant of SiO$_{2}$, $T_{ox}$:
thickness of SiO$_{2}$. Using $T_{s}$=5.0 nm at $V_{G}$=15 V, 
the carrier density
amounts to $N$=9.3$\times$10$^{17}$ cm$^{-3}$. Considering the mobility
($\mu$=0.21 $cm^{2}V^{-1}s^{-1}$) at $T_{s}$=5.0 nm, the conductivity of
pentacene is 3.1$\times$10$^{-2}$ $\Omega ^{-1}$cm$^{-1}$ for the 5.0 nm
thick film. This value is obtained on the basis of FET
characteristics. On the other hand, the conductivity can be calculated
from $I_{SD}$ (6.9$\times$10$^{-7}$ A) for the 5.0 nm thick film at
$V_{G}$=0 V, taking account of the channel width and the channel
length. This yields the conductivity of 2.6$\times$10$^{-2}$ $\Omega ^{-1}$cm$^{-1}$. The fair
agreement between these two values indicates the validity of the above
estimation, showing that the depletion layer of the organic TFT can be
explained in terms of the conventional model applicable for inorganic
semiconductors. 
Clear $V_{G}$ dependence of $d_{th}$ was observed also for
other organic films, 6T and C$_{60}$. 
In case of 6T, $d_{th}$ is 5.0 nm at
$V_{G}$=15 V, while the onset was not observed even for the 100 nm thick
C$_{60}$ film at $V_{G}$=-15 V. The thickness of the depletion layer is
inversely proportional to the major carrier concentration of the
semiconductors. Therefore, the onset was not observed for C$_{60}$ with
a low carrier concentration. Here, we comment the carrier concentration
of the present specimens. We used as-received specimens without extra
purification, and thus the impurity (carrier) concentration is rather
high, resulting in a small on/off ratio of Fig.~\ref{fig2}. However, the
high carrier concentration never affects the present estimation of
depletion layer, instead helps observe the depletion layer as a delay of
the onset at moderate condition (5 nm at $V_{G}$=15 V), and assure the
uniformity of electric field in the thin film thickness region.
 
\begin{figure}
\begin{center}
\leavevmode\epsfysize=50mm \epsfbox{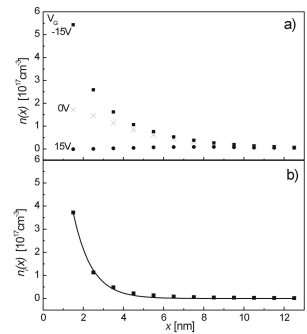}
\caption{a) Carrier density ($n(x)$) as a function of the distance ($x$)
 from the interface at various gate voltages ($V_{G}$); 15(circle),
 0(cross), and -15 V(square). b) The difference ($n_{i}(x)$)  between
 charge carrier density at $V_{G}$=-15 V and that at $V_{G}$=0 V.}

\label{fig4}
\end{center}
\end{figure}

In the previous section, we have discussed the thickness of the
depletion layer. We now proceed to the accumulation layer, relating
closely with the electric-field control of physical properties through
charge injection. Here, we assume electric fields at the interface are
determined only by $V_{G}$, independently of film thickness, and hence
the differentiation ($dI_{SD}(x)/dx$) 
corresponds to the local conductance at the
distance $x$ from the interface. The carrier density ($n(x)$) can be
approximated by dividing the local conductance by the mobility
$\mu$($x$) (see Fig.~\ref{fig3}) under the condition that the mobility
is constant in the whole film at each film thickness. The carrier
density thus obtained is shown in Fig.~\ref{fig4} as a function of the
distance $x$ from the interface. The large carrier density at $V_{G}$=0
V in small $x$ region is due to the charge transfer from the Au
electrodes to the pentacene molecules. The density of the carriers
injected by electric fields $n_{i}(x)$ is, thus, 
the difference between the
carrier density $n_{-15}(x)$ at $V_{G}$=-15 V and $n_{0}(x)$ at
$V_{G}$=0 V. As seen in the figure, $n_{i}(x)$ decays steeply with
increasing $x$, meaning that the injected carriers are localized at the
interface. The thickness of the accumulation layer is much smaller than
that of the depletion layer, which is estimated to be 5.0 nm in the
previous section. For quantitative estimation on the thickness of the
accumulation layer, $n_{i}(x)$ is fitted with an exponential function
($a\times exp(-x/b)$). The
fitted value of $b$=0.91 nm can be considered as an effective thickness
of the accumulation layer, which is just thinner than 1 ML. In other
words, most of carriers are localized in the first ML next to the
interface.

The obtained results on the accumulation and depletion layers can be a guide to the operation principle of organic TFT and electric-field control of physical properties by field induced charge injection. Since charge carries reside in the first one ML from the interface, the physical properties only at the interface could be controlled by the charge injection. This implies the importance of a well-ordered interface through which charges are efficiently injected. Furthermore, we propose that the semiconductor-metal or -superconductor transition is a promising target to study. When the semiconductor-metal transition occurs by charge injection, metallic and semiconducting regions sit side by side with only an atomic distance apart in the organic film. Under this situation, free carries can interact with exciton, associated with the semiconductor, at the interface, thus leading a possible ground for superconductivity by exciton mechanism\cite{9,10}.

\section{CONCLUSIONS}
\label{sec4}

In conclusion, we present the in-situ and continuous measurement of the
conductivity of growing organic films, as a simple but powerful method
to determine the distribution curve of injected carriers and the
dimension of the accumulation and depletion layers.

\acknowledgments{
We thank Prof. H. Aoki for discussion. The authors express their thanks for financial support from Grant-in-Aid from the Ministry of Education, Culture, Sports, Science and Technology of Japan (14GS0207).}

\end{multicols}

\begin{references}
\bibitem{4} The thickness of the depletion layer can be estimated indirectly by the capacitance measurement assuming the dielectric constant of semiconductors, but the thickness of the accumulation layer is hard to be determined.
\end{references}
\end{document}